\documentstyle[epsf]{article}

\def\bc{\begin{center}}
\def\ec{\end{center}}
\def\be{\begin{eqnarray}}
\def\ee{\end{eqnarray}}
\def\bq{\begin{equation}}
\def\eq{\end{equation}}
\def\ben{\begin{enumerate}}\def\een{\end{enumerate}}

\def\roughly#1{\mathrel{\raise.3ex\hbox{$#1$\kern-.75em
\lower1ex\hbox{$\sim$}}}}

\begin{document}

\def\bra{\langle }
\def\ket{\rangle }

\bc
\Large{{\bf 
Generalized Parton Distributions
and the structure of the
constituent quark}
\footnote{Supported in part by HPRN-CT-2000-00130, 
SEUI-BFM2001-0262 and by MIUR through the funds COFIN01}
}
\ec

\bigskip

\bc
\large Sergio Scopetta$^{a,b}$, Vicente Vento$^{c,d}$
\ec

\medskip

\bc
$^a$ Dipartimento di Fisica, Universit\`a degli Studi di
Perugia, I-06100 Perugia, Italy\\  
$^b$ Istituto Nazionale di Fisica Nucleare, Sezione di Perugia\\
$^c$
Departament de Fisica Te\`orica,
Universitat de Val\`encia, 46100 Burjassot (Val\`encia), Spain\\
$^d$
IFIC,
Consejo Superior de Investigaciones Cient\'{\i}ficas
\ec

\medskip

\noindent
In a scenario where 
the constituent quarks are composite systems,
Generalized Parton Distributions
(GPDs) are built 
from wave functions to be evaluated
in a Constituent Quark Model (CQM), 
convoluted with the GPDs of the constituent quarks
themselves. 
The approach permits to access the full kinematical
range
corresponding
to the DGLAP and ERBL regions, so that cross sections
for deeply virtual Compton scattering can be estimated.

\bigskip




\section{Introduction}

Generalized Parton Distributions (GPDs) \cite{first} 
parametrize the non-perturbative hadron structure
in hard exclusive 
processes \cite{jig}.
The measurement of
GPDs 
would represent 
a unique way to access 
crucial features of hadron structure
\cite{radnew,ji1}. 
Relevant experimental efforts to measure GPDs,
by exclusive electron Deep Inelastic Scattering
(DIS), will 
take place soon and
it is urgent to produce 
predictions for these quantities.
Recently,
calculations have been performed 
in Constituent Quark Models (CQM) \cite{epj,bpt}.
The CQM has a long story of successful
predictions in low-energy studies of the
structure of the nucleon.
At high energies,
in order to compare predictions
with data,
one has to evolve, according to perturbative QCD, the leading twist
component of the structure functions obtained
at the low-momentum scale, the ``hadronic scale''
$\mu_0^2$, associated with the model.
Such a procedure, already addressed in \cite{pape}, 
has proven
successful in describing the gross features of 
standard Parton Distribution Functions (PDFs) 
(see, e.g., \cite{trvv}).
Similar expectations motivated the study of GPDs
in Ref. \cite{epj}, where
a simple formalism has been proposed to calculate
the quark contribution to 
GPDs.
The procedure of Ref. \cite{epj} has been extended
and generalized in Ref. \cite{prd}.
As a matter of fact, the approach of Ref. \cite{epj}, when
applied in the 
forward case, has been proven to be able to
reproduce the gross features of PDFs
\cite{trvv} but,
in order to achieve a better agreement with data, it
has to be improved.
In a series of papers, it has been shown that
DIS data
are consistent with a low energy scenario
dominated by composite constituent quarks
of the nucleon
\cite{scopetta1}, defined trhough a scheme suggested
by Altarelli,
Cabibbo, Maiani and Petronzio (ACMP) \cite{acmp},
properly updated.
The same idea has been recently applied to show
the evidence of complex objects inside the proton
\cite{psr}, analyzing data of electron scattering off 
the proton.
In this talk, we review the main findings of Ref. \cite{prd},
where the same idea has been applied to obtain
realistic
predictions for GPDs and, at the same time,
to explore kinematical regions, not accessible with
the approach of Ref. \cite{epj}.
In particular, the evaluation
of the sea quark contribution becomes possible, so that
GPDs can be calculated in their full range of definition.
Such an achievement will permit to estimate
the relevant cross-sections,
providing us with a
tool for planning 
future experiments.

\section{GPDs in a constituent quark scenario}

We are interested in hard exclusive processes, such as
Deeply Virtual Compton Scattering (DVCS).
We use the formalism of Ref. \cite{ji1}.
Let us think now to a nucleon target, with initial
and final momenta $P$ and $P'$, respectively,
with $\Delta=P^\prime -P$
being the momentum transfer.
The main quantity
we discuss is
the GPD $H(x,\xi,\Delta^2)$.
In a system of coordinates where
the virtual photon 4-momentum, $q^\mu=(q_0,\vec q)$, and $\bar P=(P+P')^\mu/2$ 
are collinear along $z$,
the $\xi$ variable, the so called ``skewedness'', parameterizing
the asymmetry of the process, is 
$\xi = - n \cdot \Delta$,
where $n$
is a light-like 4-vector
with $n \cdot \bar P = 1$.
With this choice,
GPDs describe the amplitude for finding a quark with momentum fraction
$~~x+\xi/2$ (in the Infinite Momentum Frame) in a nucleon 
with momentum $(1+\xi/2) \bar P$
and replacing it back into
the nucleon with a momentum transfer $\Delta^\mu$.
Besides, when the quark longitudinal momentum fraction 
$x$ of the average nucleon momentum $\bar P$
is less than $-\xi/2$ (the so-called
negative DGLAP region), GPDs describe antiquarks;
when it is larger than $\xi/2$
(the positive DGLAP region), they describe quarks;
when it is between $-\xi/2$ and $\xi/2$ (the ERBL region), 
they describe $q \bar q$ pairs.
One should keep in mind that, besides the variables
$x,\xi$ and $\Delta^2$, GPDs depend,
as the standard PDFs, on the momentum scale $Q^2$ at which they are
measured. For an easy presentation,
such a dependence will be 
omitted.
\begin{figure}[h]
\vspace{3.3cm}
\includegraphics{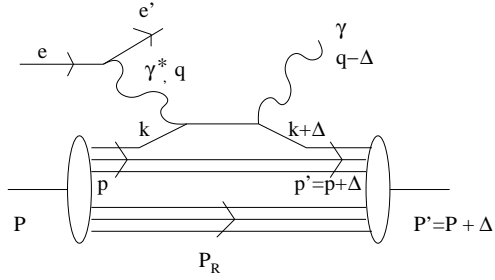}
\caption{The DVCS process
in the present approach.}
\end{figure}
In \cite{epj},
the Impulse Approximation (IA) expression
for 
$H_q(x,\xi,\Delta^2)$, suitable to perform CQM
calculations, has been obtained.
In particular, it has been shown that,
taking into account the quarks degrees of freedom only,
considering a process with $\vec \Delta^2 \ll M^2$, 
in a non relativistic framework
one has
\begin{eqnarray}
H_q(x,\xi,\Delta^2) & = & \int d \vec k\,\,\,
\delta \left (x + { \xi \over 2} - {k^+ \over \bar P^+}  \right )\,
\tilde n_q(\vec k , \vec k + \vec \Delta)
\nonumber
\\
& = & \int d \vec k\,\,\,
\delta \left (x + { \xi \over 2} - {k^+ \over \bar P^+}  \right )\,
\int e^{i ((\vec k + \vec \Delta) \vec r
-\vec k \vec r' )} \rho_q(\vec r, \vec r') d \vec r
d \vec r'\,,
\label{hnk}
\end{eqnarray}
where the one-body non diagonal charge density
$
\rho(\vec r, \vec r')$ 
and the one-body non-diagonal momentum distribution
$\tilde n_q(\vec k , \vec k + \vec \Delta)$ have been introduced.
The above equation 
allows the calculation of 
$H_q(x,\xi,\Delta^2)$ in any CQM, and
it naturally verifies the two crucial constraints
of the GPDs \cite{epj}.
With respect to Eq. (\ref{hnk}),
a few caveats are necessary.

i) If use is made of a CQM, containing only
constituent quarks 
only the DGLAP region can be explored.
In order to introduce the study of the ERBL region, so that
observables can be calculated,
the model has to be enriched.

ii) 
in Eq. (\ref{hnk}), 
the $x$ variable for the valence quarks
is not defined in its natural support.
Several prescriptions have been proposed in the past
to overcome such a difficulty in the standard PDFs case \cite{pape,trvv}.

iii) The Constituent Quarks are assumed to be point-like.

The two issues i) and ii) will be now discussed,
by relaxing the condition iii) and allowing for a 
composite structure of the constituent quark

The procedure described in the previous section,
when applied in the standard
forward case, has been proven to be able to
reproduce the gross features of PDFs
\cite{trvv}.
In order to achieve a better agreement with data, the approach
has to be improved.
In a series of previous papers, it has been shown that
DIS data
are consistent with a low-energy simple picture of the constituent quark as a
complex system of point-like partons.
\cite{scopetta1}, updating
the ACMP scenario 
\cite{acmp}. 

Following the same idea,
we describe here
a model for the reaction mechanism of an off-forward process,
such as DVCS.
As a result, a convolution formula giving
the proton $H_q$ GPD in terms of a constituent quark off-forward momentum 
distribution,
$H_{q_0}$,
and of a GPD of the constituent quark $q_0$ itself, 
$H_{q_0q}$,
will be derived.
It is assumed that the hard scattering with the virtual photon
takes place on a parton of a spin $1/2$ target, made of
complex constituents. 
The scenario we are thinking to is depicted in Fig. 1 for the 
special case of DVCS.
In addition to the kinematical variables,
$x$ and $\xi$, already defined,
one needs few more ones to describe the process.
In particular,
$x'$ and $\xi'$, for the ``internal''
target, i.e., the constituent quark, have to be introduced
\cite{prd}.
Using IA,
a standard procedure 
can applied
and a  convolution formula,
is readily obtained (see Ref. \cite{prd} for details):
\begin{eqnarray}
H_{q}(x,\xi,\Delta^2) \simeq  
\sum_{q_0} \int_x^1 { dz \over z}
H_{q_0}(z, \xi ,\Delta^2 ) 
H_{q_0 q}\left( {x \over z},
{\xi \over z},\Delta^2 \right ) 
\label{main}
\end{eqnarray}
where 
$H_{q_0}(z, \xi ,\Delta^2 )$
is to be evaluated in a 
given CQM, according to Eq. (\ref{hnk}),
for $q_0=u_0$
or $d_0$, while $H_{q_0 q}( {x \over z},
{\xi \over z},\Delta^2)$ is the constituent quark
GPD, which is still to be modelled.
We can start modelling this quantity
thinking first of all to its forward limit, 
where ``constituent quark
parton distributions'' have to be recovered.
As we said,
in a series of papers
\cite{scopetta1} 
a simple picture of the constituent quark as a
complex system of point-like partons
has been proposed, 
re-taking the ACMP scenario, \cite{acmp}.

Let us recall the main features of that idea.
The constituent quarks are
complex objects whose structure functions are described by a set of functions
$\phi_{q_0q}(x)$ that specify the number of point-like partons 
of type $q$ which
are present in the constituent of type $q_0$, with fraction $x$ of its total
momentum. We will call these functions  the structure
functions of the constituent quark.
The functions describing the nucleon parton distributions are expressed in
terms of the independent $\phi_{q_0q}(x)$ and of the constituent 
density distributions ($q_0=u_0,d_0$) as,
\bq
q(x,Q^2)=\sum_{q_0}\int_x^1 {dz\over z} 
q_0(z,Q^2) \phi_{q_0q} \left ({x \over z},Q^2 \right)~,
\label{fx}
\eq
where $q$ labels the various partons, i.e., valence quarks  
($ u_v,d_v$), sea quarks ($u_s,d_s, s$), sea antiquarks ($\bar u,\bar d, \
\bar
s$) and gluons $g$.
The different types and functional forms of the structure functions of the 
constituent quarks are derived from three very natural assumptions 
\cite{acmp}:
i) The point-like partons are $QCD$ degrees o freedom, 
i.e. quarks, 
antiquarks and gluons;
ii) Regge behavior for $x\rightarrow 0$ and duality ideas;
iii) invariance under charge conjugation and isospin.
The last assumption 
of the approach relates to the 
choice of the scale
at which the constituent quark 
structure is defined. We choose 
$\mu_0^2$ = 0.34 GeV$^2$ \cite{trvv,grv}. This hypothesis fixes $all$ 
but one the parameters of
the approach.
The remaining paramter is fixed according to 
the value of $F_2$ at low x\cite{acmp},
and its value is chosen again according to \cite{grv}.
The values of the parameters obtained 
are listed in \cite{scopetta1}.
\begin{figure}[h]
\vspace{5.3cm}
\includegraphics{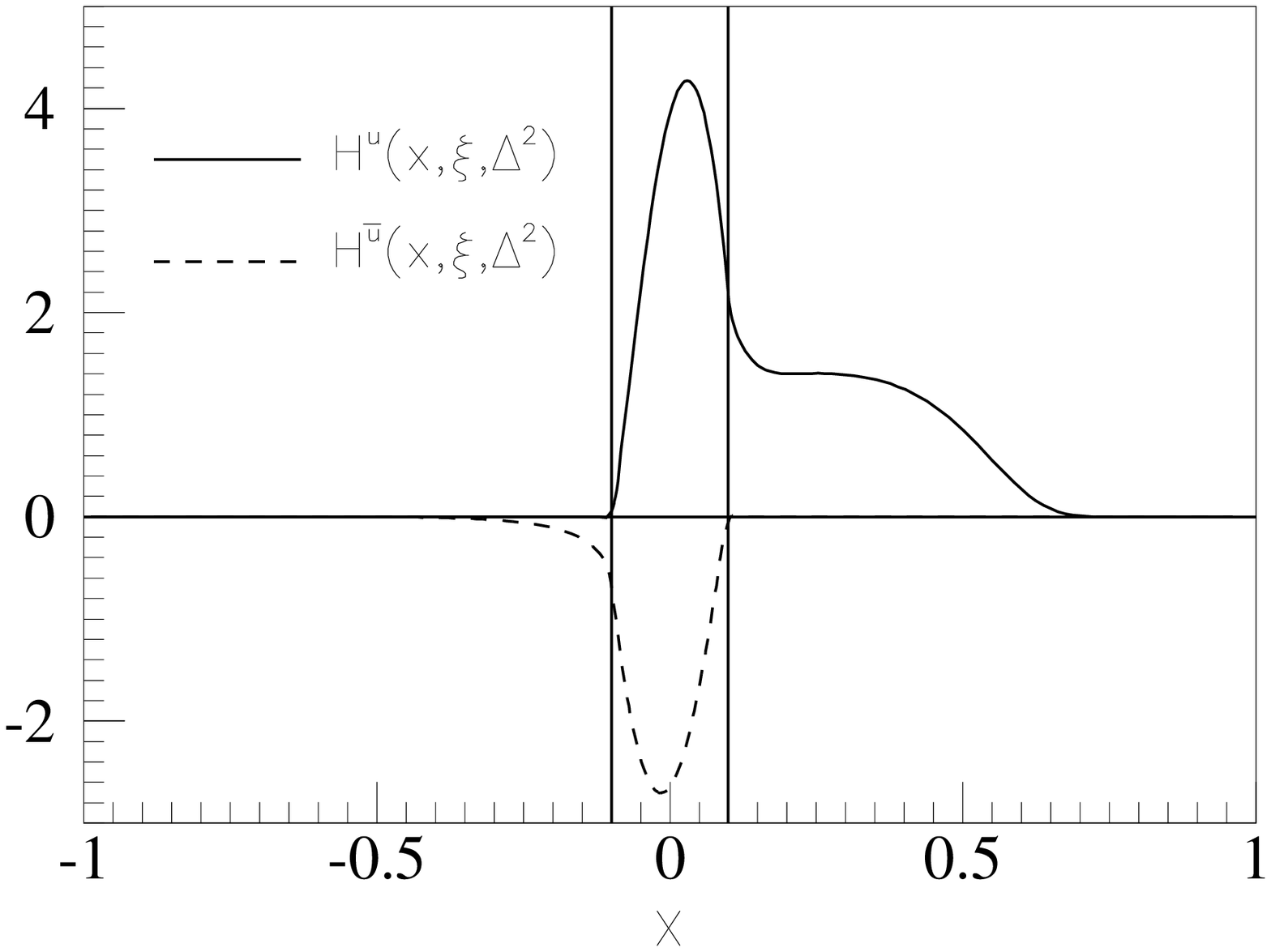}
\includegraphics{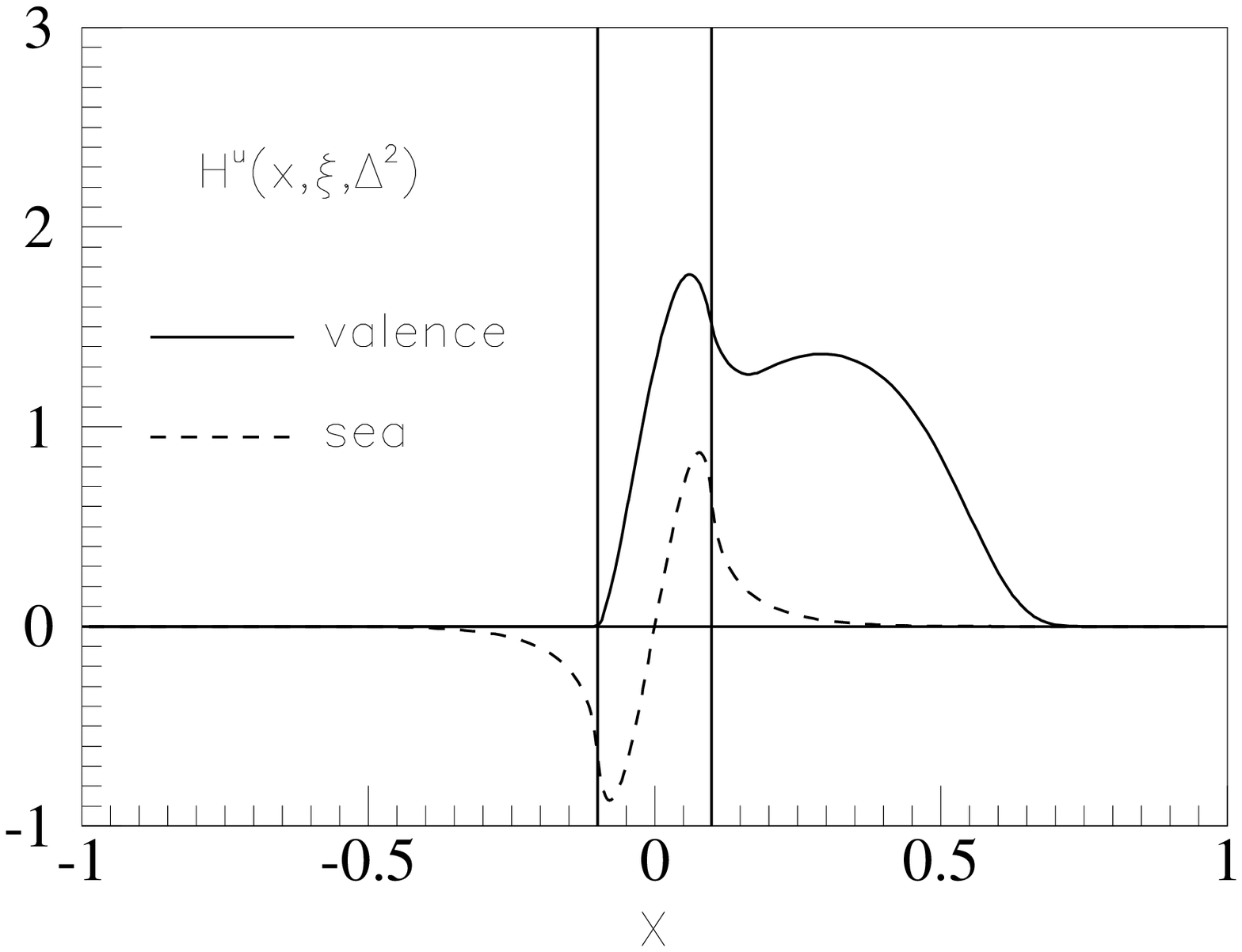}
\caption{Left: the $u$-quark (full) and $u$-antiquark (dashed) $H$.
Right: the $u$ valence (full) and sea 
(sea-quarks plus antiquarks)(dashed) $H$ GPD.
Results are shown 
for $\xi=0.2$ and $\Delta^2 = -0.5$ GeV$^2$, 
at the scale of the
model.
}
\end{figure}
The other ingredients appearing in Eq. ({\ref{fx}}), i.e., 
the density distributions for 
each constituent quark, are defined according to 
Eq. (\ref{hnk}). Now we have to generalize this scenario 
to describe off-forward 
phenomena. Of course, the forward limit of our
GPDs formula, Eq. (\ref{main}), has to be given by 
Eq. ({\ref{fx}}):
this gives
$H_{q_0 q}( {x \over z},
{\xi \over z},\Delta^2)$ \cite{prd}.

Now the off-forward behaviour of the Constituent Quark GPDs 
has to be modelled.
This can be done in a natural way by using the ``$\alpha$-Double
Distributions'' (DD's) language proposed by Radyushkin \cite{radd}.
DD's, $\Phi(\tilde x, \alpha, \Delta^2)$,
represent an alternative parametrization of the matrix elements
describing DVCS and hard exclusive electroproduction processes, 
with respect to the one based on GPDs.
The DD's do not depend on the skewedness parameter $\xi$; rather,
they describe how the {\sl{total}}, $P$, and {\sl{transfer}}, $\Delta$,
momenta are shared between the interacting and final partons, by means
of the variables $\tilde x$ and $\alpha$, respectively.
$H_{q_0q}$ for the constituent quark target,
is related to the $\alpha$-DD's, which we call
$\tilde \Phi_{q_0 q} (\tilde x, \alpha,\Delta^2)$ for the constituent
quark,
in the following way:
\begin{eqnarray}
H_{q_0q}(x,\xi,\Delta^2) = \int_{-1}^1 d\tilde x
\int_{-1 + |\tilde x|}^{1-|\tilde x|} 
\delta \left(\tilde x + {\xi \over 2} \alpha - x \right)
\tilde \Phi_{q_0 q} (\tilde x, \alpha,\Delta^2) d \alpha~.
\label{hdd}
\end{eqnarray}

In \cite{radd}, a factorized
ansatz is suggested for the DD's: 
\begin{equation}
\tilde \Phi_{q_0 q} (\tilde x, \alpha,\Delta^2) =
h_{q} (\tilde x, \alpha,\Delta^2)
\Phi_{q_0 q} (\tilde x) 
F_{q_0}(\Delta^2)
\label{ans}
\end{equation}
where
the $\alpha$ dependent
term, 
$h_{q} (\tilde x, \alpha,\Delta^2)$, 
has the character of a mesonic amplitude,
$
\Phi_{q_0 q} (\tilde x) 
$
represents the forward density
and, eventually, 
$
F_{q_0}(\Delta^2)
$
the constituent quark f.f.
In the following, we will assume for the constituent quark GPD the 
above factorized form, so that we need to model the three functions
appearing in Eq. (\ref{ans}), according to the description
of the reaction mechanism we have in mind.
For the amplitude $h_q$, use will be made of
one of the simple normalized forms suggested in
\cite{radd}.
Besides, in our approach the forward densities
$\Phi_{q_0q} (\tilde x)$ have to be given by
the standard $\Phi$ functions of the 
ACMP approach (see Ref. \cite{prd} for details).
\begin{figure}[h]
\vspace{6.cm}
\includegraphics{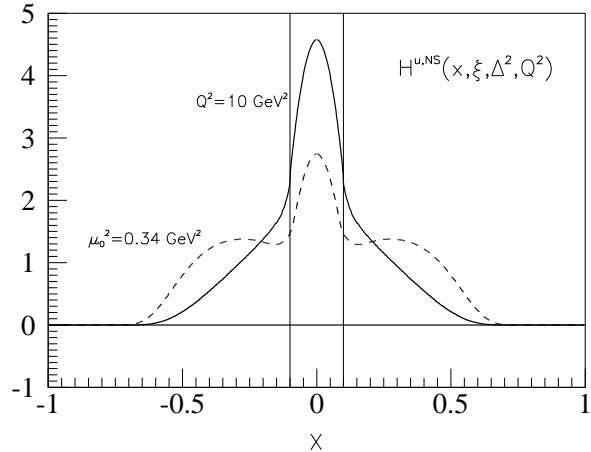}
\caption{
The $NS$ $H$ GPD for valence $u$-quarks
at $\xi=0.2$ and $\Delta^2 = -0.5$ GeV$^2$, at the momentum scale of the
model, $\mu_0^2=0.34$ GeV$^2$ (dashed), and after NLO-QCD evolution
up to $Q^2=10$ GeV$^2$ (full).
}
\end{figure}
Eventually, as a f.f. we will take a monopole form
corresponding to a constituent quark size $r_Q \simeq 0.3 fm$,
supported by the analysis of \cite{psr}.
By using these ingredients,
Eq. (4) can be calculated and 
the result, when cast into Eq. (\ref{main}), yields the
GPD in our ACMP approach.

Results for the GPD $H_q$ of the proton have been obtained calculating
Eq. (\ref{main}). 
$H_{q_0q}$, has been
calculated according to the model discussed above, and 
$H_{q_0}$ has been evaluated according to Eq. (\ref{hnk}),
using, as an illustration, the wave functions
of the Isgur and Karl (IK) model \cite{ik}. 
A prescription introduced in \cite{prd}
has been used to correct the poor-support
problem, addressed in Section 2.

Results are shown in Fig. 2.
A relevant tail of the valence quarks
contribution in the ERBL region is found,
in agreement with other estimates \cite{jig}.
The knowledge
of GPDs in the ERBL region is necessary for
the calculation of cross-sections.
The ERBL region is accessed here, with
respect to the calculation of Ref. \cite{epj}, thanks
to the constituent structure which has been introduced.

The results shown so far have to be related to
the hadronic scale $\mu_0^2$=0.34 GeV$^2$.
In Fig. 3 the NLO QCD-evolution 
of the Non Singlet, valence $u$ distribution,
is shown.
In evolving the results,
the approach of Ref. \cite{evo} has been applied
and a code provided by A. Freund has been used.
The evolution clearly shows a strong enhancement of the ERBL region. 

\section{Conclusions}

The aim of the talk has been to describe composite
constituent quarks \cite{acmp,scopetta1}
in studies of GPDs \cite{prd}.
This is the continuation of an effort to construct a scheme 
which describes hadrons
in different kinematical and dynamical scenarios. 
We have developed a formalism which expresses the hadronic GPDs 
in terms of constituent quarks GPDs by means of appropriate
convolutions. 
Looking at our results, we discovered that such a scheme transforms 
a hadronic model, in whose original description only
valence quarks appear, into one containing
all kinds of partons.
Moreover, the starting model produces no structure in
the ERBL region, while after the structure of the constituent quark 
has been incorporated, it does. The completeness
of the $x$-range,
for the allowed
$\Delta^2$ and $\xi$, of the present description,
permits the calculation of cross-sections in a wide
kinematical range and it can be used therefore
to guide future experiments.

\end{document}